\documentclass[twocolumn,amsmath,amssymb,floatfix]{revtex4} 
\usepackage{epsfig}
\usepackage{amsmath,amssymb}
\newcommand{\bma}{\begin{math}}
\newcommand{\ema}{\end{math}}
\newcommand{\beq}{\begin{equation}}
\newcommand{\eeq}{\end{equation}}
\newcommand{\bc}{\begin{center}}
\newcommand{\ec}{\end{center}}

\def\n{{\bf  \hat n}}
\def\k{{\bf k}}

\begin{document}

\title{Probing the Friedmann equation during recombination with future CMB experiments}

\author{Oliver Zahn$^{1,2,3}$\footnote{Email address:
    zahn@mpa-garching.mpg.de} and Matias
    Zaldarriaga$^{3,4}$\footnote{Email address: mz31@nyu.edu}}

\address{$^1$Fakult\"at f\"ur Physik, Ludwig-Maximilians-Universit\"at, Geschwister-Scholl-Platz 1, 80799 M\"unchen}

\address{$^2$Max-Planck-Institut f\"ur Astrophysik, P.O. Box 1317,
  85741 Garching}

\address{$^3$Dept. of Physics, New York University, 4 Washington Pl., New York, NY 10003}

\address{$^4$Institute for Advanced Study, Einstein Drive, Princeton
  NJ 08540 \\ {\rm(Received 16 December 2002; revised manuscript received 19
  February 2003; published 18 March 2003)}}

\begin{abstract}

We show that by combining measurements of the temperature and polarization anisotropies of the Cosmic Microwave Background (CMB), future experiments will tightly constrain the expansion rate of the universe during recombination. A change in the expansion rate modifies the way in which the recombination of hydrogen proceeds, altering the shape of the acoustic peaks and the level of CMB polarization. The proposed  test is similar in spirit to the examination of abundances of light elements produced during Big Bang Nucleosynthesis and it constitutes a way to study possible departures from standard recombination. For simplicity we parametrize the change in the Friedmann equation by changing the gravitational constant $G$. The main effect on the temperature power spectrum is a change in the degree of damping of the acoustic peaks on small angular scales. The effect can be compensated by a change in the shape of the primordial power spectrum. We show that this degeneracy between the expansion rate and the primordial spectrum can be broken by measuring CMB polarization. In particular we show that the MAP satellite could obtain a constraint for the expansion rate $H$ during recombination of $\delta H/H \simeq 0.09$ or $\delta G/G \simeq 0.18$ after observing for four years, whereas Planck could obtain $\delta H/H \leq 0.014$ or $\delta G/G \leq 0.028$ within two years, even after allowing for further freedom in the shape of the power spectrum of primordial fluctuations.

\end{abstract}

\maketitle

\section{Introduction\label{introduction}}

The parameters of our cosmological model will be determined with great
accuracy by upcoming data from Cosmic Microwave Background
experiments, galaxy surveys, weak lensing surveys, Lyman alpha forest
studies, and other observations. With the new data it will be possible to perform a number of consistency checks that will strengthen our confidence in
the underlying model. Some of these consistency checks have already
been performed with existing data.  Recent analysis of the CMB data
have resulted in constraints on the baryon density $\Omega_b h^2$ that
are in excellent agreement with its determination based on the study
of the primordial abundances of light elements
(e.g. \cite{Wang,Efsthathiou}). The combination of CMB data with local
measures of the Hubble constant \cite{keyproject} and measures of the
local strength of galaxy clustering result in a determination of the
cosmological constant that is in good agreement with results from the
study of the luminosity of distant supernovae
(e.g. \cite{Riess,revpar}). Recently also joint BBN and CMB
constraints on different dark energy models have been discussed in \cite{Kneller:2002zh}.

One of the assumptions of the cosmological model that has been hard to
test is the explicit validity of the Friedmann equation -- the
relation between the expansion rate of the universe and its matter
content. The difficulty lies in finding an epoch in the evolution of
the universe during which both the energy density and the expansion
rate can be determined independently. The two obvious candidates are
the present time and Big Bang Nucleosynthesis (BBN).

Precise measures of both the expansion rate and the matter density in
the local universe are difficult to accomplish and suffer from various
systematic problems. At present the expansion rate has been measured
with errors on the order of $10\%$ \cite{keyproject}. Direct
determinations of the matter density however are more uncertain. It is fair to say that the general conclusion
from these studies is that the Friedmann equation only holds true if
either a cosmological constant or a curvature term is added. This is
because direct determinations of the present matter density almost
always point to $\Omega_m < 1$ (e.g. see \cite{revpar}).  Neither the
cosmological constant nor the curvature scale can be constrained
independently, that is without going through the Friedmann equation, so
at best we can say that the Friedmann equation has not been tested
accurately at the present epoch. A more radical interpretation would
be that the Friedmann equation has been tested but that the test has
failed. Although we do not support this interpretation, at present
some infrared modification of gravity cannot be ruled out
observationally. Such modifications are being
explored for example as ways to solve the cosmological constant
problem (see for example \cite{Arkani-Hamed}) or to explain the
accelerated expansion inferred from the luminosity distance to high
redshift supernovae without resorting in a cosmological constant or a
quintessence field (e.g. see \cite{Deffayet:2001pu,Deffayet2}).

During the epoch of Big Bang Nucleosynthesis (BBN) the situation is
more fortunate. The energy density is dominated by radiation which we
think we can estimate accurately. On the other hand the expansion rate
affects the freeze-out abundances of light elements, so that a
precision test of the Friedmann equation can be performed.  The
standard procedure is to constrain the number of relativistic degrees
of freedom, $g_*$, which can be translated into a limit on the number
of neutrino species. A lot of progress has been made in determining
the primordial abundances. Recently the deuterium abundance in
hydrogen clouds at high redshift was accurately determined
\cite{O'Meara:2000dh,Burles:1997ez,Burles:1997fa}. Building on a prior
of $N_\nu \geq 3$ these data have been exploited to enforce an upper
limit of 3.2 at $2\sigma$ for the number of neutrino species, based
purely on BBN considerations \cite{Burles:1999zt}.

Progress has also been made in appreciating the systematical
uncertainties that impair the determination of primordial $^4He$ (see
e.g. \cite{Olive:2000qm,Peimbert:2002ks}). Built on the safely
established abundance ranges for Deuterium, Helium and Lithium, it can be shown that the uncertainty in the number of relativistic degrees of freedom during BBN is around 9 \% (68\% C.L.)\cite{Olive:1998vj}.
This constraint can equally well be phrased as a constraint on the
validity of the Friedmann equation: during Nucleosynthesis the ratio
of the squared expansion rate to the energy density can depart by only
$9 \%$ from what is predicted by the Friedmann equation. Constraints on the expansion history during BBN have also been established in \cite{Carroll:2001bv}.

In this paper we propose using the anisotropies in the CMB to perform
a test similar to the one that has been done using BBN.  We will show
that such a test can ultimately constrain the validity of the
Friedmann equation during recombination more accurately than what has
so far been reached within Nucleosynthesis, albeit in a more model
dependent way.

During recombination, the energy density is dominated by the density
of non-relativistic matter which we cannot estimate directly. However
the dark matter energy density enters in two different ways and one
can exploit this to simultaneously determine the dark matter density
and the expansion rate during recombination.

The ratio of matter to radiation energy density sets the redshift of
matter radiation equality. At that time the expansion rate changes
from a scaling as $t^{1/2}$ to $t^{2/3}$. Perturbation modes of the
photon-baryon fluid that entered the horizon during the radiation
dominated era behave differently than those that entered during the
matter dominated era. Modes that entered during radiation domination
provided the dominant contribution to the total density perturbation
that generated the gravitational potential. On the contrary modes that
entered the horizon during matter domination, were sub-dominant in
their contribution to the total density perturbation which was
dominated by the dark matter fluctuations. The gravitational potential
acts as a source for perturbations in the photon baryon fluid. As a
result, small scale modes that entered the horizon in the radiation
era go through a sort of feedback loop that increases their amplitude
as they cross the horizon (for a review of CMB physics see for example
\cite{DodlesonHu}). The anisotropy power
spectrum is very sensitive to the redshift of matter radiation
equality and thus the CMB should very accurately determine the ratio
of dark matter to radiation energy density, i.e. the parameter
$\Omega_m h^2$.

The dark matter density dominates over other energy components in the
Friedmann equation during recombination. In the standard scenario, it
sets both the redshift at which matter and radiation become equal and
the rate of expansion during recombination through the Friedmann
equation. In this paper we break the link between energy density and
expansion rate by introducing a free parameter to modify the Friedmann
equation. We investigate how well such a parameter can be constrained.

From a pragmatic perspective our study can be regarded as the
investigation of a particular departure from standard
recombination. Different variations have been studied in the
literature. For instance the possibility that energetic sources of
Ly-$\alpha$-photons could be present during recombination to delay it
was considered in \cite{Peebles:200l}. Also, the possiblility of a
time variation of the fine structure constant was investigated
\cite{Kaplinghat:1998ry,Landau:2000dd,Hannestad:2000ys}. In
\cite{Uzan:2002vq} the effects of a time dependence of the
gravitational constant have been outlined. The conclusion of these
investigations and ours is that with future CMB data departures from
standard recombination will be severely constrained and perhaps
modifications that point to interesting new physics could be
discovered.

In section \ref{lambda}, we will introduce our model and in section
\ref{analysis} we will investigate the constraints that can be set with
currently available data and forecast what future CMB experiments
might be able to achieve. We will conclude in section \ref{discussion}
with discussion.

\section{The Model: Variation of the gravitational constant\label{lambda}}

In this section we introduce the model we will use to investigate how
well one can test the Friedmann equation using the CMB.  The
problem is somewhat more subtle than in the case of BBN because we are
dealing with the dynamics of perturbations that could be affected by the 
``new physics''  in ways other than through a change in
the expansion rate. Thus we need to find a self-consistent way of
modifying both the dynamics of the universe and that of the
perturbations.

One possibility is to add another component that contributes to the
energy density during recombination, increasing the rate of expansion
at that time. The additional component could be a quintessence field
with a potential and initial conditions tuned so that it has some
effect during recombination and is unimportant or only marginally
important at other times (except perhaps today when it could start to
dominate). This approach has the virtue of only modifying the
expansion rate but it introduces too much freedom because results
depend on when exactly this extra component is important. In such a 
model we also expect the ratio of the sound horizon at recombination
to the angular diameter distance to the last scattering surface to
change and thus that the acoustic peaks be slightly shifted. At late
times the evolution of the gravitational potentials will also induce
an integrated Sachs-Wolfe (ISW) effect. As a result, constraints on
any specific model of this kind will come from these three effects
\cite{UzanRiaz}.

In our study we want to isolate the information encoded in the change
of the expansion rate at recombination so we will use a simpler
prescription and assume that the gravitational constant $G$ is
somewhat different from its locally measured value. We introduce a
single parameter $\lambda$ such that,
\begin{equation}
G \rightarrow \lambda^2 G.
\end{equation}
The expansion rate is proportional to $\lambda$.  With this
prescription not only the Friedmann equation gets modified but also
the dynamics of the perturbations changes because it depends on the
strength of gravity. We will show in the following that our
prescription has the nice feature that it only changes the CMB power
spectrum through the change in recombination, allowing us to isolate
the observable effects of this change. The basic reason is that
gravity does not have a preferred scale and that we only measure
angles when studying the CMB. If $G$ were slightly different all that
would happen is that the universe would be expanding a bit faster or
slower by a factor $\lambda$ so that the ``expansion clock" would be
running at a different rate. Such a change cancels in the ratios of
distances that we measure with the CMB. The only way we can find out
that such an alteration had occurred is by having an independent clock
that measures the expansion rate. In our case this independent clock
will be the physics of hydrogen recombination.  In this sense our
simple test is very similar to what has been done in the context of
BBN.

\subsection{Effect of $\lambda$ on the CMB anisotropies} 

The dependence of the Hubble parameter and the dynamics of
perturbations on the gravitational constant will lead to modifications
of the CMB anisotropies as we vary the parameter $\lambda$. We will
discuss the physics in this section.

We start by considering the modification to the Friedmann equation,
\begin{equation}
H^2=\left(\frac{\dot{a}}{a} \right)^2=\frac{8 \pi}{3} G \rho \rightarrow \frac{8 \pi}{3} \lambda^2 G  \rho
\end{equation}
where $\rho$ is the total energy density.  As a function of the expansion factor $a$ and $\lambda$, the expansion rate $H$ satisfies:
\begin{equation}
H(a,\lambda)=\lambda f(a) \ ,
\end{equation}
where the function $f(a)$ is independent of $\lambda$.
Thus with this simple prescription, the shape of the function $H$ of $a$ is not changed by $\lambda$, only the amplitude changes. For example the redshift at which matter and radiation contribute equally to the energy density does not change. The change introduced is a simple rescaling of the ``expansion rate clock".
 
In order to understand how the anisotropies get modified, we start by writing down the integral solution for the temperature anisotropies produced by a mode of wavevector $\k$ observed towards direction $\n$ \cite{Seljak:1996is}. The  temperature can be written as an integral along the line of sight over sources,
\begin{equation}
\Delta T(\n,\k) = \int^{\tau_0}_0  d \tau \;  S(k,\tau) e^{i \k\cdot \n D(\tau)}  g(\tau)
\label{lineofsight}
\end{equation}
In this equation $S(k,\tau)$ is the source term,  $g(\tau)$ is the visibility function, and $D(\tau)$ is the distance from the observer to a point along the line of sight corresponding to the conformal time $\tau$ ($ad\tau=dt$). Dots indicate differentiation with respect to $\tau$. 

The visibility function $g(\tau)$ can be written in terms of the opacity for Thomson scattering $\kappa$ as 
\begin{equation}
g(\tau)=\dot{\kappa} \exp(-\kappa)=-d/d\tau \exp(-\kappa)
\end{equation}
with
\begin{equation}
\kappa = \sigma_T \int_\tau^{\tau_0} a n_e(\tau) d\tau,
\label{opacity}
\end{equation}
where $\sigma_T$ is the Thomson scattering cross section and $n_e(\tau)$ is the number density of free electrons. We have also defined $\dot{\kappa}=\sigma_T a n_e $. Finally, the source term in the integral equation is given by
\begin{equation}
S=\phi + \frac{\delta_\gamma}{4} + \hat{n} \cdot \mathbf{v}_b
\end{equation}
where $\phi$ is the gravitational potential, $\delta_\gamma$ is the fractional perturbation in the photon energy density and $\mathbf{v}_b$ is the baryon velocity.  

The acoustic oscillations in the photon-baryon plasma satisfy (see e.g. \cite{Hu:1994jd}) 
\begin{eqnarray}
\ddot{\delta}_\gamma & +& \frac{\dot{R}}{(1+R)}\dot{\delta}_\gamma + k^2 c_S^2\delta_\gamma = \nonumber\\
&=& 4\left[\ddot{\phi} + \frac{\dot{R}}{1+R}\dot{\phi} -\frac{1}{3} k^2 \phi \right] \;
\label{photbarglg}
\end{eqnarray}
with the sound speed $c_S^2=1/3(1+R)$, and the baryon-photon momentum density ratio $R = (p_b+\rho_b)/(p_\gamma + \rho_\gamma)\simeq  3\rho_b/4 \rho_\gamma$. The velocity satisfies,
\begin{equation}
\dot{\delta}_\gamma+kv_\gamma+\dot{\phi}=0
\end{equation}
Finally the gravitational potential satisfies the Poisson equation
\begin{equation}
-k^2 \phi = 4 \pi \lambda^2  G \rho \delta^{total} ,
\label{po}
\end{equation}
where $\rho \delta^{total}$ gives the combined perturbation due to all the fluids.

We are now ready to study the dependence of $\Delta T$ on $\lambda$. For this purpose it is best to consider the expansion factor as a time variable rather than $\tau$. We note that,
\begin{equation}
\frac{d}{d \tau}=\frac{da}{d \tau}\frac{d}{da} = a^2 \cdot H\cdot \frac{d}{da}=\lambda f(a) a^2  \frac{d}{da}.
\label{tautoa}\end{equation}
As a result, when we change time variables, every time derivative introduces a factor of $\lambda$. By inspection of equations (\ref{photbarglg}) and  (\ref{po}) it is clear that the dynamics of a mode with wavenumber $k$ in a universe with $\lambda \neq 1$ is equivalent to the dynamics of a mode with $k'=k/\lambda$ in a universe with $\lambda=1$. That is, 
\begin{equation}
S(k,a,\lambda)=S(k/\lambda ,a, \lambda=1).
\end{equation}
We have explicitly included the $\lambda$ dependence of the source to make our argument clearer. 
 
To obtain the CMB power spectrum, we first need to expand equation (\ref{lineofsight}) in Legendre polynomials.  The amplitude of the $l$ expansion coefficient is 
\begin{equation}
\Delta T_l(k,\lambda) = \int^{1}_0  da \; \tilde S(k,a,\lambda) j_l(k D(a,\lambda)) \tilde g(a,\lambda).   \label{lineofsight2}
\end{equation}
We have introduced $\tilde g(a,\lambda) = -d/da \exp(-\kappa)$. The conformal distance $D$ is given by
\begin{equation}
D(a,\lambda) =\int_{a}^1 \frac{da}{H(a) a^2}=\lambda^{-1} D(a,\lambda=1) .
\end{equation}
Thus if the visibility function where to be independent of $\lambda$ we would have,
\begin{equation}
\Delta T_l(k,\lambda) = \Delta T_l(k/\lambda,\lambda=1).
\label{lineofsight3}
\end{equation}
The power spectrum is calculated from $\Delta T_l(k,\lambda)$ using,
\begin{eqnarray} 
C_l(\lambda) &=& \int \frac{dk}{k} P(k) |\Delta_{Tl}(k,\lambda)|^2 \nonumber \\
&=&  \int \frac{dk'}{k'} P(k' \lambda) |\Delta_{Tl}(k',\lambda=1)|^2,
\end{eqnarray} 
where $P(k)$ is the power spectrum of primordial fluctuations which is usually taken to be a power law $P(k)\propto k^{n-1}$.  Thus we see that provided we adjust the amplitude of the primordial power spectrum appropriately $C_l(\lambda)=C_l(\lambda=1)$. 

Our result is qualitatively very easy to understand: gravity introduces no preferred scale, so the dynamics of the perturbations remains the same when scales are measured in units of the expansion time. As a result, the angular power spectrum does not change as we change $\lambda$. 

Of course this conclusion only holds true if the visibility function
is not affected by $\lambda$. However the physics of recombination
does introduce a preferred timescale, so the power spectra of the
anisotropies will actually change. In other words, in our simple
minded prescription the only source of change is the difference in the
way recombination proceeds as we change the expansion rate of the
universe at recombination. This is the sense in which our model
resembles the studies done in the context of Big Bang Nucleosynthesis.

\begin{figure}
\begin{center}
\epsfig{file=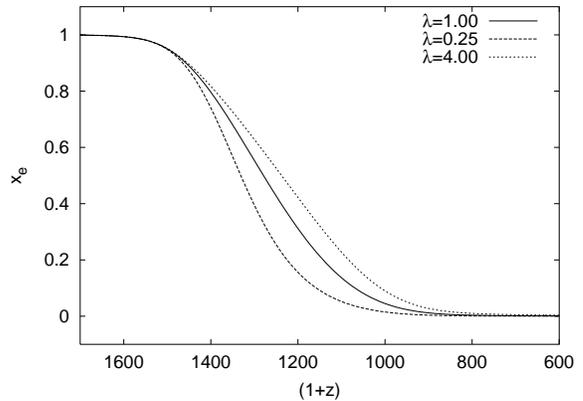,width=5.5cm,angle=-90}
\caption{Ionization fraction as a function of redshift for three values of $\lambda=0.25,1,4$.\label{freeeminus}}
\end{center}
\end{figure}

Let us now turn to study how the visibility function changes with $\lambda$. It depends on the ionization fraction $x_e = n_e/n_H$, where $n_e$ again is the free electron density and $n_H$ is the number density of hydrogen atoms. The evolution of the ionization fraction is modified when $G$ is changed. It evolves according to (e.g. \cite{Ma:1994ub}):
\begin{equation}
\frac{dx_e}{d \tau} = a C_r \left[\beta(T_b)(1-x_e)-n_H\alpha^{(2)}(T_b)x_e^2\right]
\label{recombdgl}
\end{equation}
where $a(t)$ is just the scale factor, $\beta(T_b)$ is the collisional
ionization rate from the ground state and $\alpha^{(2)}(T_b)$ is the
recombination rate to excited states. The baryon temperature is $T_b$
and the Peebles correction coefficient (which also depends on the
expansion rate) is denoted $C_r$ \cite{Peebles:1968}.  The
transformation from $\tau$ to $a(\tau)$ as a time variable using
equation (\ref{tautoa}), makes clear, that contrary to what happens to
the perturbation equations, $x_e(a)$ depends on $\lambda$. We plotted
$x_e$ for different values of $\lambda$ in Figure
\ref{freeeminus}. The behavior is easy to understand; the faster the
universe is expanding at a given redshift (i.e. the larger the
$\lambda$), the more difficult it is for hydrogen to recombine and
hence the larger is $x_e$.

\begin{figure}
\begin{center}
\epsfig{file=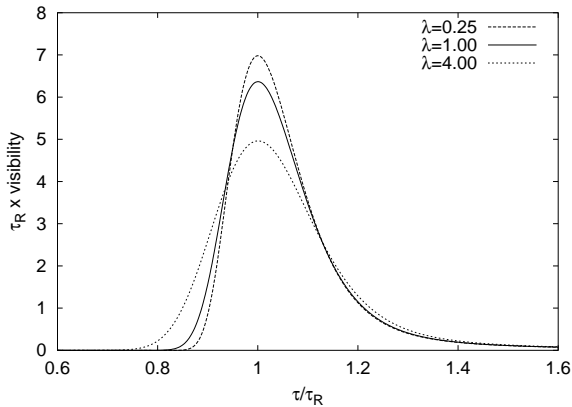,width=5.5cm,angle=-90} 
\caption{The visibility function as a function of conformal time for $\lambda=0.25,1,4$. The axes have been rescaled to take out the overall scaling of $\tau$ with $\lambda$.\label{visibility}}
\end{center}
\end{figure}

\begin{figure*}
\begin{center}
\epsfig{file=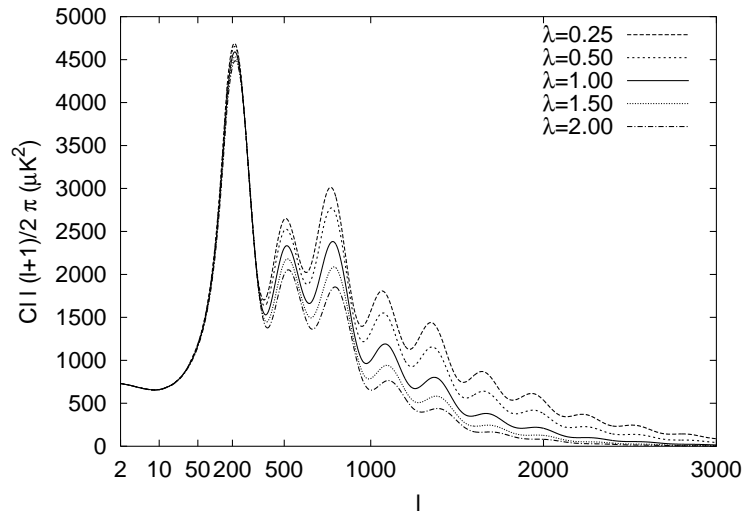, width=7cm,angle=-90}
\end{center}
\caption{Effect of $\lambda$ on the Temperature power spectrum: Higher peaks are more severely damped as $\lambda$ increases while the height of the first peak is almost unchanged.\label{cmbGplot}}
\end{figure*}

The change in $x_e$ leads to a change in the visibility function which we show in Figure \ref{visibility}. 
As $\lambda$ is increased, the visibility function becomes broader. This broadening leads to a larger damping of the anisotropies on small (angular) scales, as shown in Figure \ref{cmbGplot}. We note however that even for a factor of four change in $\lambda$ the changes in the visibility function are rather small. What happens is that if we increase $\lambda$, $x_e$ at a given redshift after the start of recombination increases. However, when calculating optical depths this change almost exactly cancels with the decrease in the time intervals between different redshifts due to the increased expansion rate. As a result changes in both the location and shape of the visibility function are small even for large changes in $\lambda$. 

Figure \ref{cmbGplot} shows that the effect of $\lambda$ is to change the relative amplitudes of the acoustic peaks on different scales. This effect can be compensated by changing the relative amplitude of modes of different scales in the primordial power spectrum. 

\begin{figure*}
\begin{center}
\epsfig{file=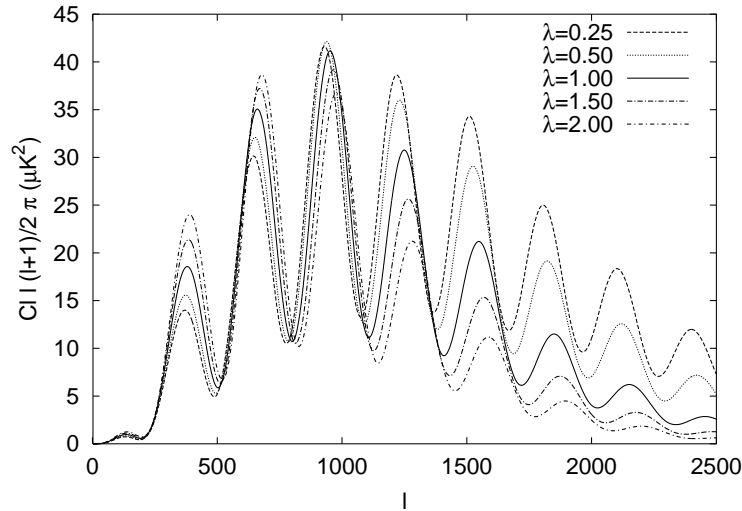,width=7cm,angle=-90}
\caption{The effect of $\lambda$ on E-type polarization power
spectra. On large angular scales the polarization signal will be
boosted, while on small scales, it will be damped.\label{ppower}}
\end{center}
\end{figure*}

\begin{figure*}
\begin{center}
\epsfig{file=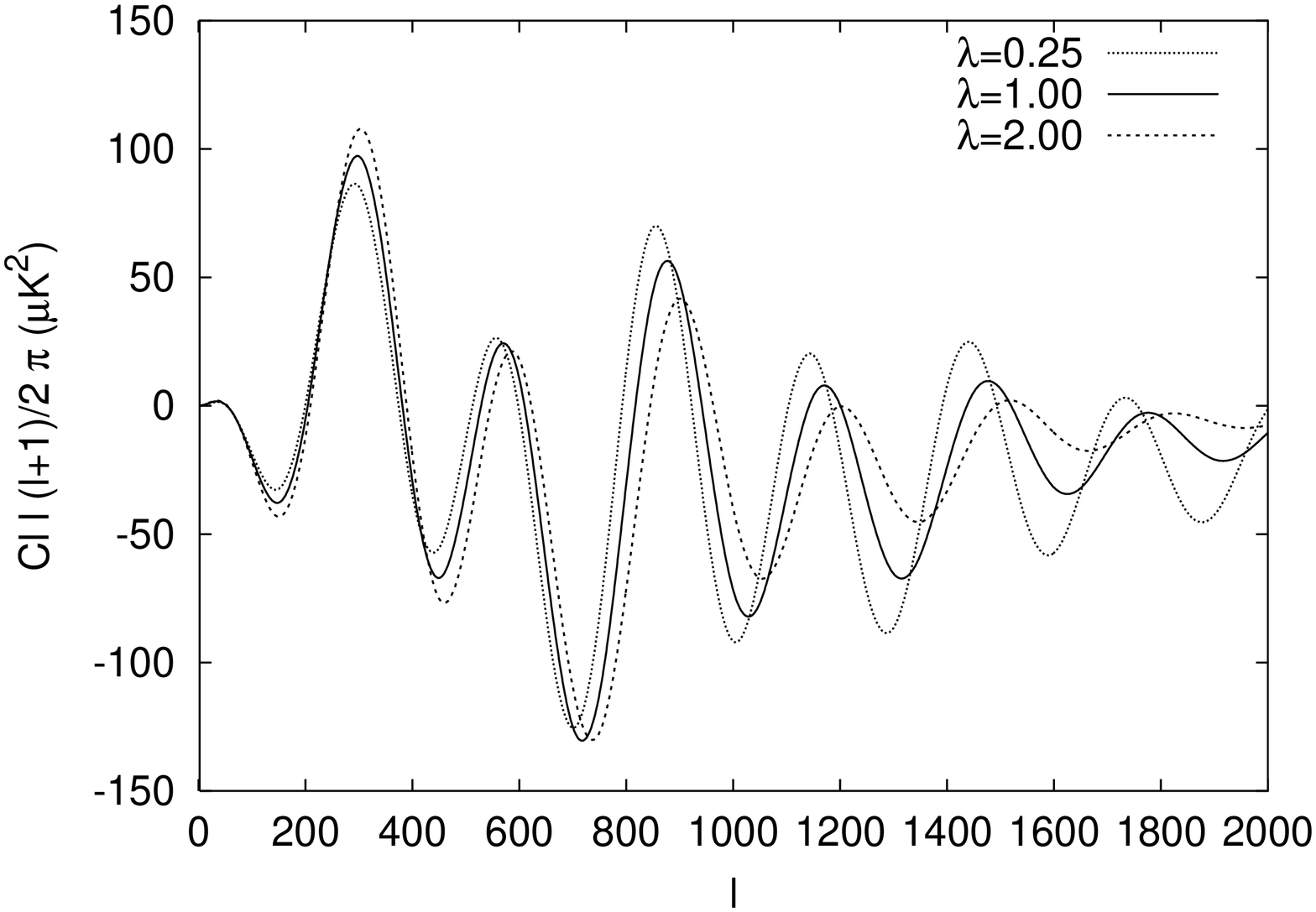,width=7cm,angle=-90}
\caption{The effect of $\lambda$ on the cross correlation between
temperature and E-type polarization.\label{cross}}
\end{center}
\end{figure*}

We are now going to study what happens to CMB polarization and to show
that it can lift the degeneracy between $\lambda$ and the shape of the
primordial power spectrum. To understand the effect of $\lambda$ on
the polarization we will employ the simple analytic expression for the
amplitude of the $Q$ Stokes parameter produced by a single Fourier
mode $k$ \cite{Zaldarriaga:1995gi}:
\begin{equation}
Q \propto c_s k \delta\tau_D \sin (k c_S \tau_D) e^{-k^2/k_D^2}
\end{equation}
where $\tau_D$ is the conformal time corresponding to the peak of the visibility function, $\delta \tau_D$ is its width and $k_D$ describes the damping of the small scale modes. As we discussed above, the damping increases with the width of the visibility function, so the exponential factor will lead to the same effect we described for the temperature. The difference in the case of the polarization is the extra $\delta \tau_D$ in the amplitude of the polarization. This extra factor comes from the fact that if the visibility function is wider the photons will travel on average longer between their last two scatterings which will enhance the quadrupole anisotropy and will thus lead to a higher level of polarization \cite{Zaldarriaga:1995gi}.  For that reason, there exists a characteristic wavemode value $k^*$, for scales larger than which the polarization power spectrum will increase with $\lambda$, while it will decrease for scales smaller than $k^*$.

The polarization power spectrum is plotted in Figure \ref{ppower}.  We see that for $l$s larger than around $800$, the polarization behaves just as the temperature does, decreasing for increasing $\lambda$. On larger scales the effect is opposite, the amplitude of polarization relative to temperature roughly increases by 10\% when $\lambda$ increases by 20\%. This response of the anisotropies on $\lambda$ is what will help to break the degeneracy between $\lambda$ and the primordial power spectrum when information from polarization is included.

For completeness we show the temperature polarization cross correlation power spectrum in Figure \ref{cross}. The cross correlation will be easier to detect in experiments such as the MAP satellite where the accuracy of the polarization measurement is limited by detector noise.  The cross correlation power spectrum behaves similarly to the polarization power spectrum; when $\lambda$ is increased the power on small scales is suppressed while on large scales it is amplified.

\section{Constraints on $\lambda$\label{analysis}}

In our likelihood analysis of currently available and simulated future data we let vary $\Omega_\Lambda$, $\omega_\text{dark matter}=\Omega_\text{dark matter} h^2$, $\omega_\text{baryon}=\Omega_\text{baryon} h^2$, the optical depth due to reionization $\tau_{ri}$, $\lambda$ and the amplitude and shape of the primordial power spectrum. We explicitly assume that the universe is flat.

As we mention above, we expect there to be a degeneracy between the shape of the primordial power spectrum and the parameter $\lambda$. To study this effect in the case of future satellite missions which will measure polarization and could break this degeneracy, we introduce additional freedom in the shape of the spectrum. Rather than just assuming that $P(k)$ is of power law form
\begin{equation}
P(k)=k^{n-1} \Leftrightarrow \frac{\ln P(k)}{\ln (k)}=n-1,
\end{equation}
we also allowed spectra with curvature by adding another term in the expansion of $\ln P$ as a function of $\ln k$. 
\begin{equation}
\ln P(k)/P(k_0) = (n-1) \ln(k/k_0) + \alpha [\ln (k/k_0)]^2+ ...
\end{equation}
where $k_0$ is the pivot point. With this  prescription the effective slope of the power spectrum changes slightly with scale, 
\begin{equation}
\frac{\partial \ln P(k)}{\partial \ln k}= n-1 + \alpha \ln (k/k_0) .
\end{equation}

We employed two kinds of likelihood analysis. For an evaluation of what currently available data can tell us about $\lambda$ we used an importance sampling Markov-chain method to generate a large number of cosmological models distributed according to the likelihood distribution $\mathcal{L}(\text{model}|\text{data})$. This is more efficient than an exploration of the entire parameter space, because the sampling is weighted and statistics can be established over the target distribution itself. Moreover the algorithm is very easy to implement. The weighted sampling is achieved through the Metropolis-Hastings algorithm. The application of the Markov method to the extraction of cosmological parameters from CMB information has been suggested in \cite{Christensen:2001gj}.

To speed up the power spectrum computations for the Markov chain we made use of the k-splitting technique discussed in \cite{Tegmark:2000qy}. For each model we marginalized analytically over the amplitude of the scalar fluctuations. In the end we constructed histograms and computed expectation values and variances for each parameter directly from the Markov chain.

In order to estimate what the satellite missions MAP and Planck will be able to tell us about the relation between the energy density and the expansion rate of the universe during recombination, we investigated the shape of the likelihood function $\mathcal{L}(C_l|\theta_i)$ (for a power spectrum $C_l$ given a model consisting of the cosmological parameters $\theta_i$) in the vicinity of its maximum directly by a Fisher matrix evaluation. This method has been widely used to make predictions for the errors that are to be expected in the extraction of cosmological parameters from planned CMB-experiments. The Fisher matrix is given by the expectation value of the second derivative of the logarithm of the likelihood function $\mathcal{L}(C_l|\theta_i)$. Assuming Gaussianity of the likelihood it is of the form
\begin{equation}
F_{ij}=\sum_l \sum_{A,B} \frac{\partial C_{Al}}{\partial
\theta_i}\mathbb{C}^{-1}(\hat{C}_{Al},\hat{C}_{Bl})\frac{\partial C_{Bl}}{\partial \theta_j} \; ,
\end{equation}
where $A$ and $B$ run over the three observables: temperature,
E-type polarization, T-E cross correlation and $i,j$ run over the
cosmological parameters.  The covariance matrix between parameters is
given by the inverse of the Fisher matrix. Overall we verified a good
agreement between Fisher matrix and Markov chain results which
confirms that the likelihood function $\mathcal{L}(C_l|\theta_i)$
resembles relatively well a Gaussian in the vicinity of its maximum
value.

\subsection{Constraints from currently available temperature data}\label{tempanalysis}

\begin{figure}
\epsfig{file=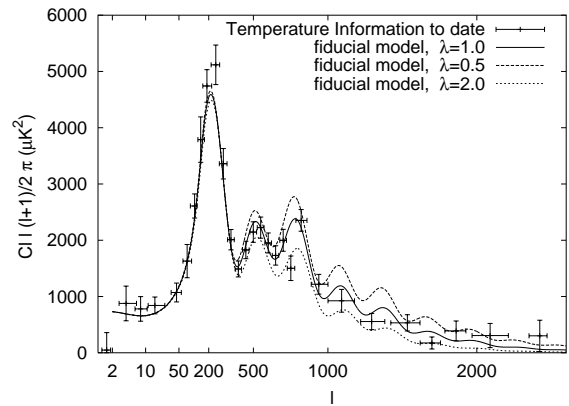,width=5.5cm,angle=-90}
\caption{Compilation of current CMB temperature anisotropy data, superimposed are three models with differing values of $\lambda$.\label{temp-data}}
\end{figure}

In order to find the constraints which can be imposed on the parameter
$\lambda$ using available temperature data, we employed a compilation
of 30 experiments which functions as a complete account of pre-MAP CMB
temperature anisotropy information \footnote{The results of the
  Archeops \cite{Benoit:2002mk} and ACBAR experiments
  \cite{Kuo:2002ua} as well as the new Boomerang data \cite{Ruhl:2002cz} were added to the compilation described in \cite{Tegmark:2002cy}.}. The data, which have been compressed to 29 bins, are plotted in Figure \ref{temp-data}. In this context, we actually assumed a simple power law behaviour of the primordial scalar perturbations. 

\begin{figure}
\begin{center}
\epsfig{file=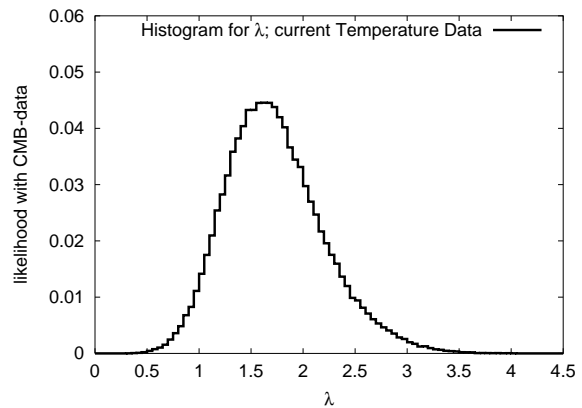,width=5.5cm,angle=-90}
\caption{Histogram for likelihood of $\lambda$ given present CMB temperature anisotropy information.\label{curthist}}
\end{center}
\end{figure}

The results we obtained from the Markov-chain analysis are shown in Table \ref{curtres} and Figure \ref{curthist}. As expected from the rather weak dependence of the anisotropies on $\lambda$  we find that current data cannot put severe constraints on the expansion rate during recombination even though other cosmological parameters are well determined.   

\begin{table}
\begin{center}
\begin{tabular}{|c|c|c|}
\hline
Param. & mean & $\sigma_\text{Markov}$  \\
\hline   
$\lambda$ & 1.749  &  0.471    \\    
\hline
$n_S$   &   1.038   &    0.0553    \\
\hline
\end{tabular}
\end{center}
\caption{Mean value and standard deviation $\sigma$ for the currently available temperature data.\label{curtres}}
\end{table}

Having emphasized the importance of measuring the linearly polarized
component of the CMB, we should also note that its recent first
detection by DASI \cite{Kovac:2002fg}, unfortunately has too large
errorbars to deliver much information about $\lambda$. Adding the DASI data in fact only changes the standard error in the determination of $\lambda$ by
3\%.

\subsection{Future satellite missions and the relevance of measuring polarization}\label{polanalysis}

An up to date estimate of the expected angular resolutions and
sensitivities of the MAP and Planck satellites has been obtained from
the experimental groups websites. For both satellites we combined
the three frequency channels with the highest angular resolution and
took into account the number of polarized instruments. In the case of
MAP the sensitivity estimate has been provided for 2 years of
observation and the angular resolutions $\theta_{\text{fwhm}}$ are 13.2,
21.0 and 31.8 arcminutes for the three channels observing at 90, 60 and 40
$GHz$. This leads to a raw sensitivity of

\begin{eqnarray*}
w_T^{-1} &=& (0.081 \mu K)^2 \\  
w_P^{-1} &=& (0.114 \mu K)^2  
\end{eqnarray*}
If MAP observes for four years the raw sensitivities ($w^{-1}$) will
be halved. For the Planck satellite mission (here the
sensitivity estimates are given for a one year observation period), the three
channels (217, 143 and 100 GHz) at $\theta_{\text{fwhm}}=$ 5.0,7.1 and
9.2 arcminutes give together
\begin{eqnarray*}
w_T^{-1} &=& (0.0084 \mu K)^2 \\
w_P^{-1} &=& (0.0200 \mu K)^2 \, .
\end{eqnarray*}
Again the raw sensitivities for a two year observation will
be  half of  these values.   For both  satellites, a  sky  coverage of
$f_{Sky}=0.8$ was assumed.

From these experimental characterisics the full estimator covariance matrices for each multipole $l$  can be constructed (e.g. \cite{Zaldarriaga:1997ch}). The diagonal terms of the covariance matrices for temperature, polarization and cross correlation are
{\small 
\begin{eqnarray}
{\rm Cov}(\hat{C}_{T l}^2)&=&\frac{2}{(2l+1) f_{Sky}} (C_{T l}+w_{T}^{-1} B_l^{-2})^2 \\
{\rm Cov}(\hat{C}_{E l}^2)&=&\frac{2}{(2l+1) f_{Sky}} (C_{E l}+w_{E}^{-1} B_l^{-2})^2 \\
{\rm Cov}(\hat{C}_{C l}^2)&=&\frac{1}{(2l+1) f_{Sky}} [C_{Cl}^2+(C_{Tl}+w_{T}^{-1}B_l^{-2})\\
  &\times& (C_{El}+w_{P}^{-1}B_l^{-2})]
\label{Clerror}
\end{eqnarray}
}where the beam window function $\mathcal{B}_l$ is to be constructed from the relevant frequency channels with their individual sensitivities $w_c$ as
\begin{eqnarray}
\mathcal{B}^2_l &=& \sum_c B_{l,c}^2 \frac{w_c}{w} \\
B_{l,c}^2 &=& e^{-l (l+1) \theta_b^2,c}
\end{eqnarray}
Here, the standard width of the beam $\theta_b$ is obtained from the full width half maximum resolution by
\begin{equation}
\theta_b,c=\frac{\theta_{\text{fwhm,c}}}{\sqrt{8 \ln 2}}
\end{equation}

As a fiducial model we have adopted the parameter values
\begin{equation}
\begin{array}{|c|c|c|c|c|c|c|c|}
\hline   
\tau   &   \Omega_K & \Omega_\Lambda & \omega_{\text{dm}} & \omega_{\text{ba}} & n_S & \lambda  \\
\hline
0.05     &     0 &  0.7   &  0.14  &  0.02  &  1  &   1   \\
\hline
\end{array}
\end{equation}
They imply a hubble constant of $h=0.73$.

\subsubsection{Expected constraints from MAP}

The analysis of expected constraints from the \emph{MAP} satellite
mission shows the constraints it can put on $\lambda$ are rather
weak. Better sensitivity is needed to accurately determine the
polarization and higher angular resolution to map the damping tail. On
the other hand our results illustrate how the inclusion of a curvature
term for the primordial spectrum significantly weakens the constraints
that one obtains from the temperature data.

A precision test of the Friedmann equation will have to wait
until both polarization and the damping tail are measured
accurately. In the near future experiments such as Boomerang, BICEP, Polatron
and others are expected to significantly improve polarization
measurements while CBI, ACBAR and others will
map the damping tail. In a few more years, the Planck satellite will
measure both temperature and polarization accurately enough to
severely constrain any change in $\lambda$. We will present an
analysis of Planck's sensitivity in the next section.

\begin{figure*}
\epsfig{file=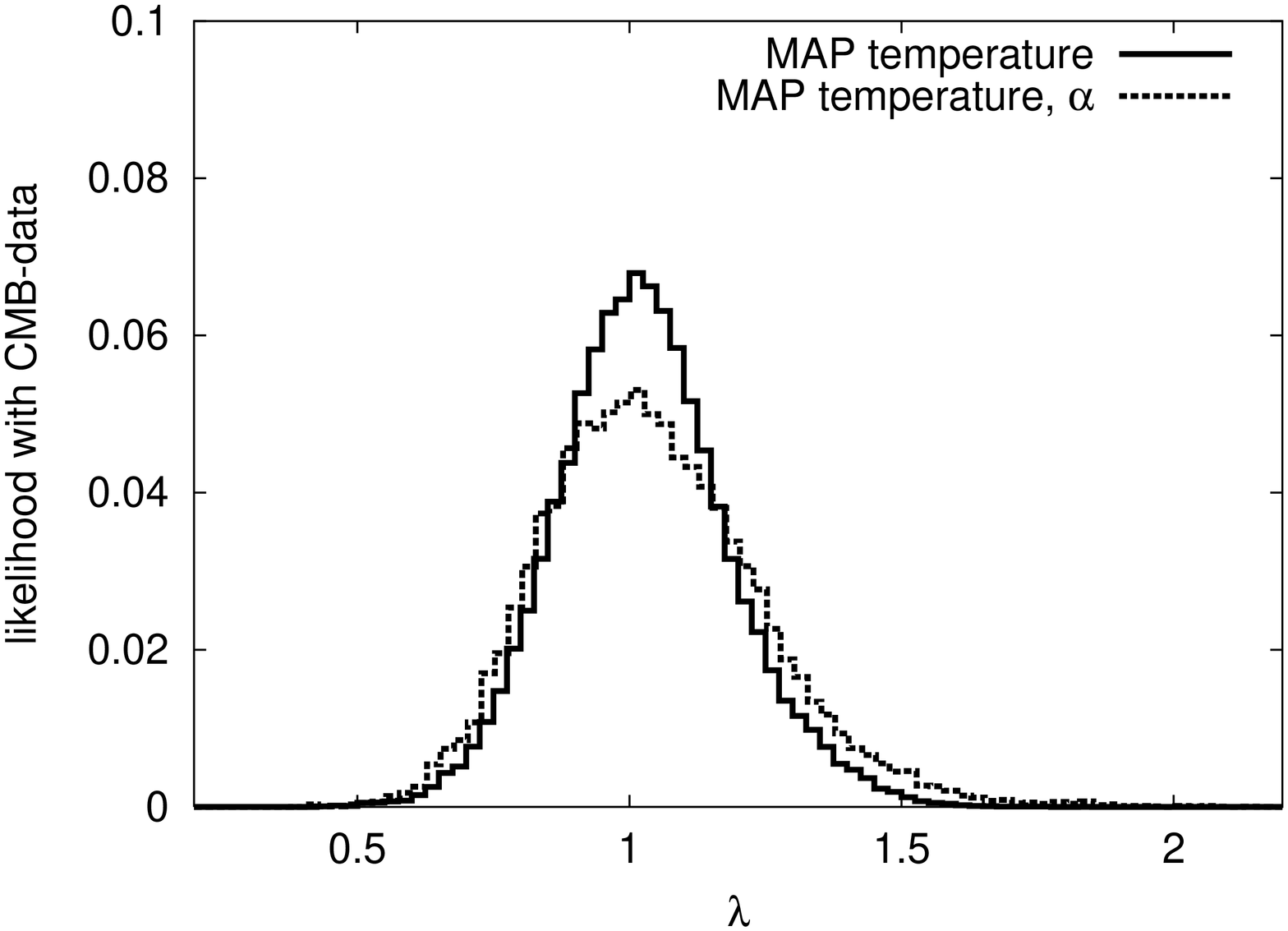,width=5cm,angle=-90}  \hspace{1cm}
\epsfig{file=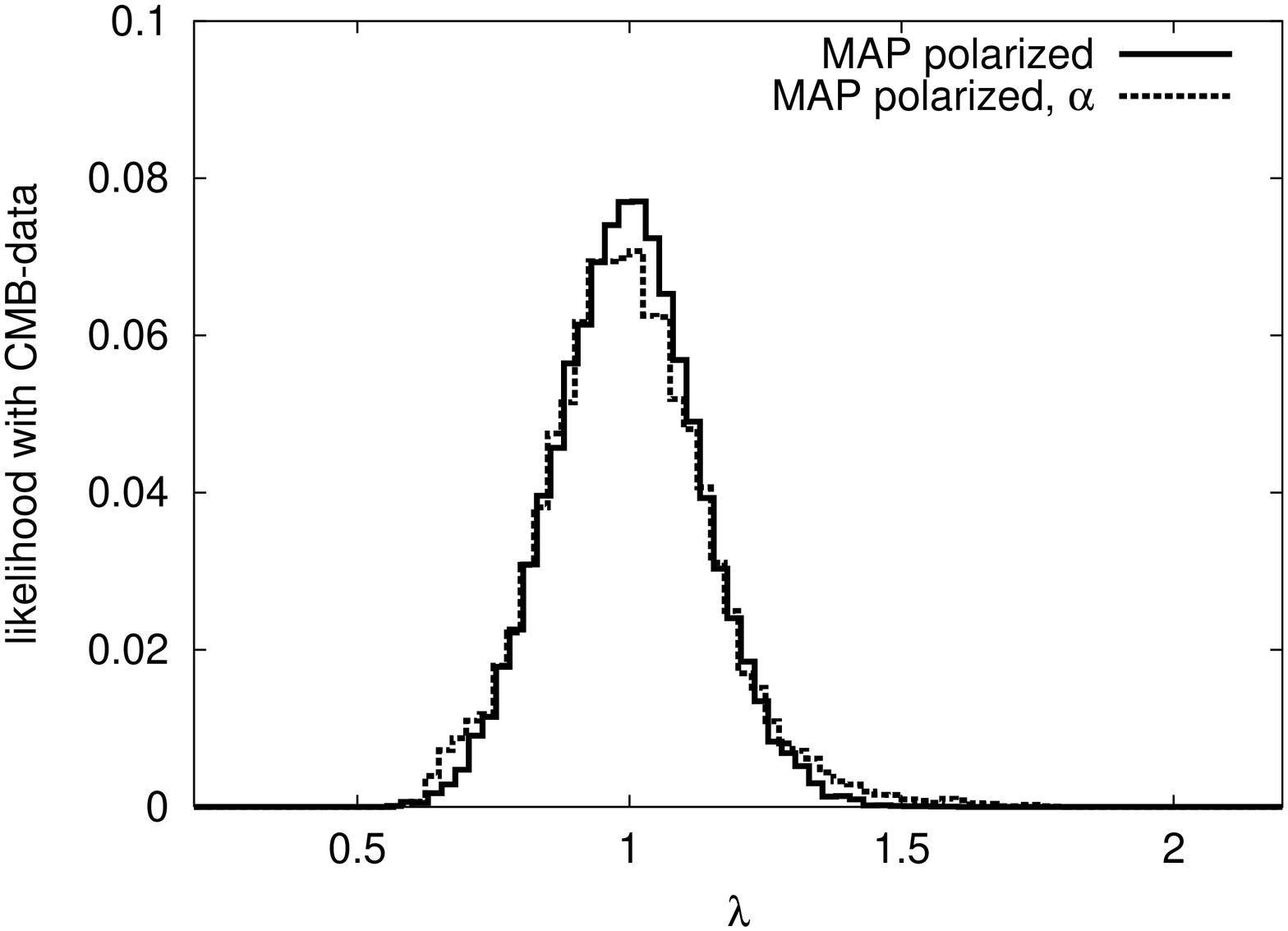,width=5cm,angle=-90}  
\caption{Likelihood distributions of $\lambda$ for a one year observation with MAP. Compared is the case in which one assumes a simple power law spectrum of the initial perturbations (solid lines) with the case in which one leaves further freedom to $P(k)$ (dashed lines). To obtain the Figure on the left, only temperature information was used and for the one on the right polarization was included.\label{maphist}}
\end{figure*}

\begin{figure*}
\epsfig{file=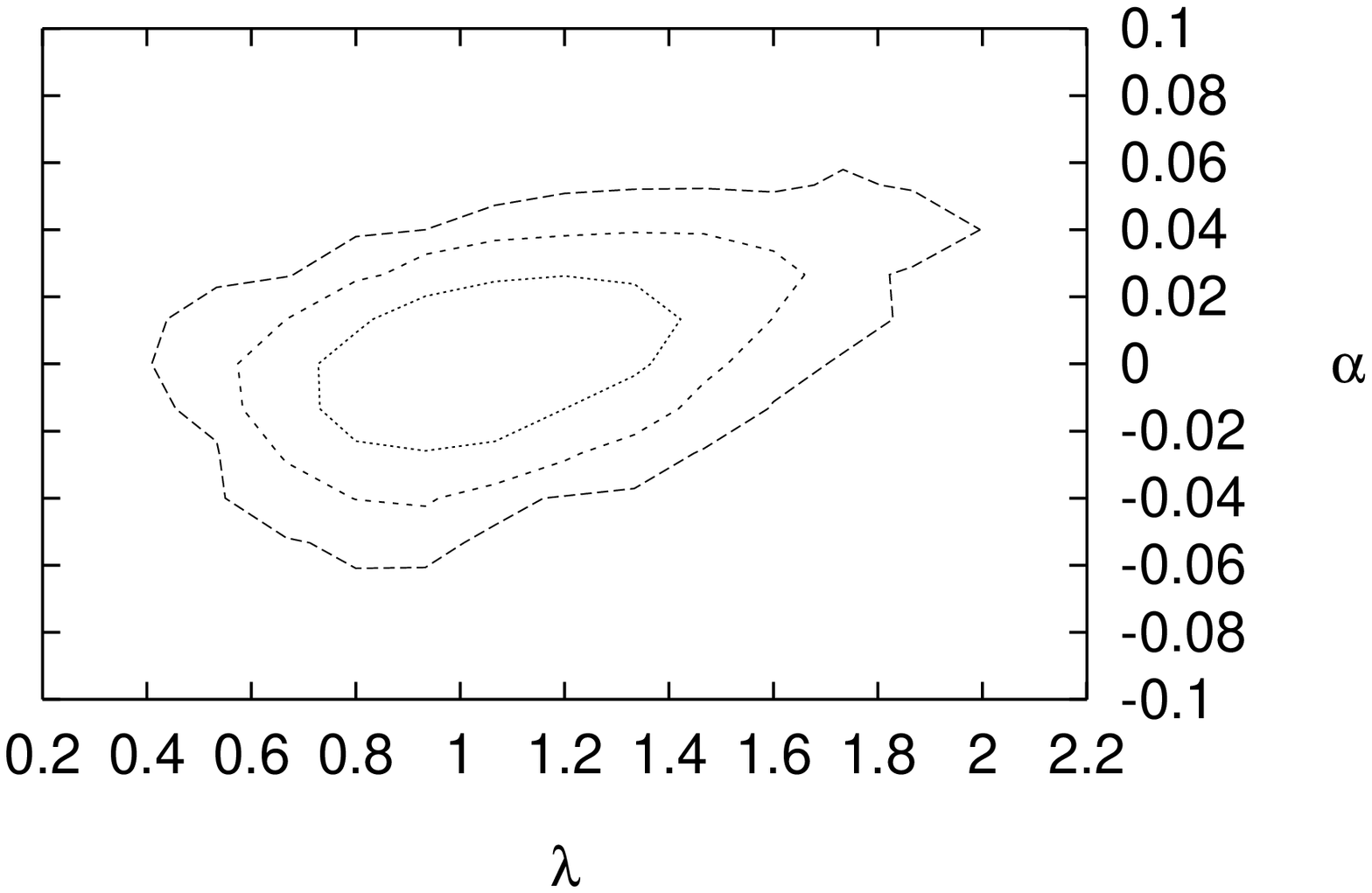,width=7cm,angle=-90} \hspace{-3cm}
\epsfig{file=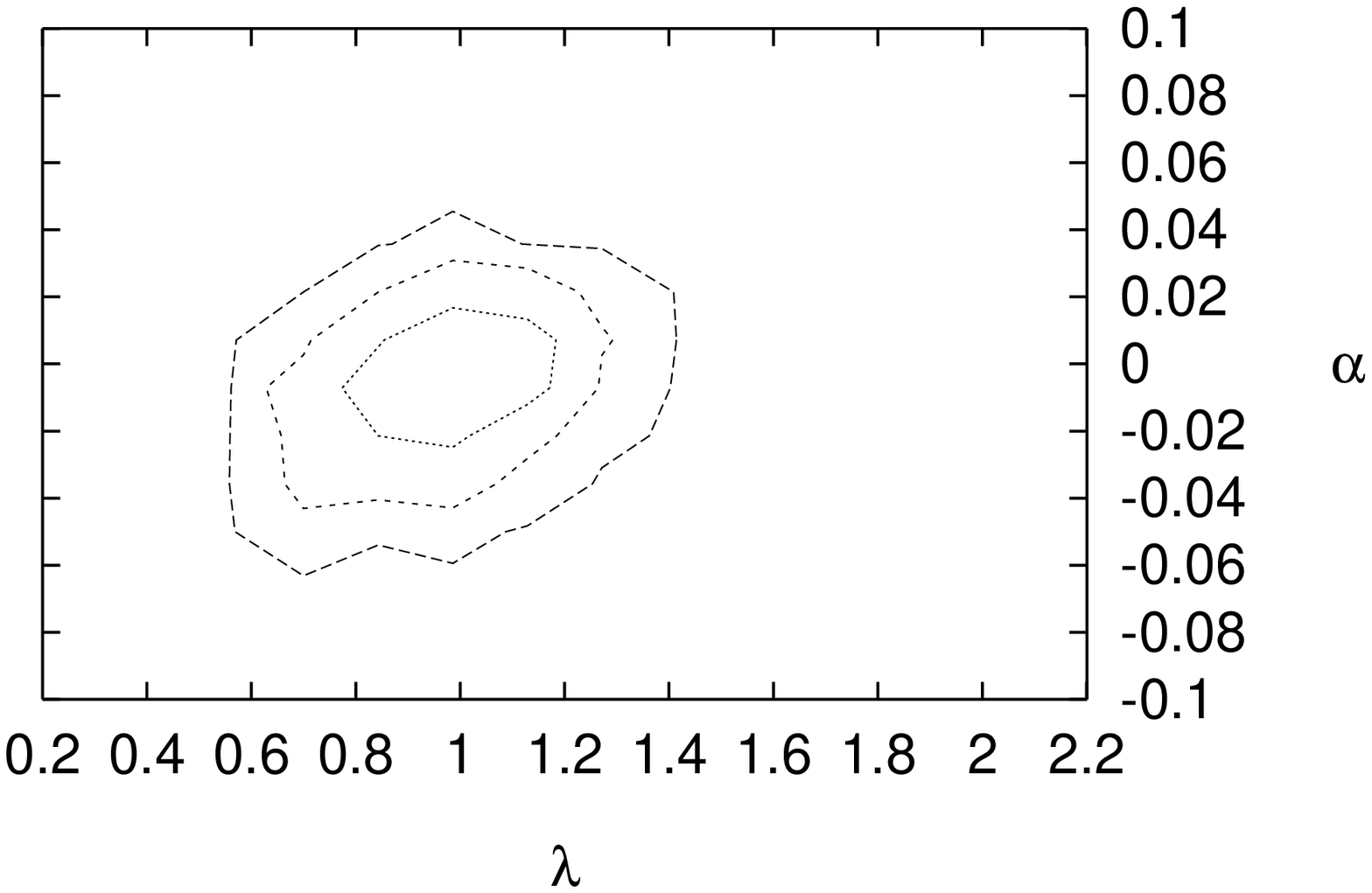,width=7cm,angle=-90}
\epsfig{file=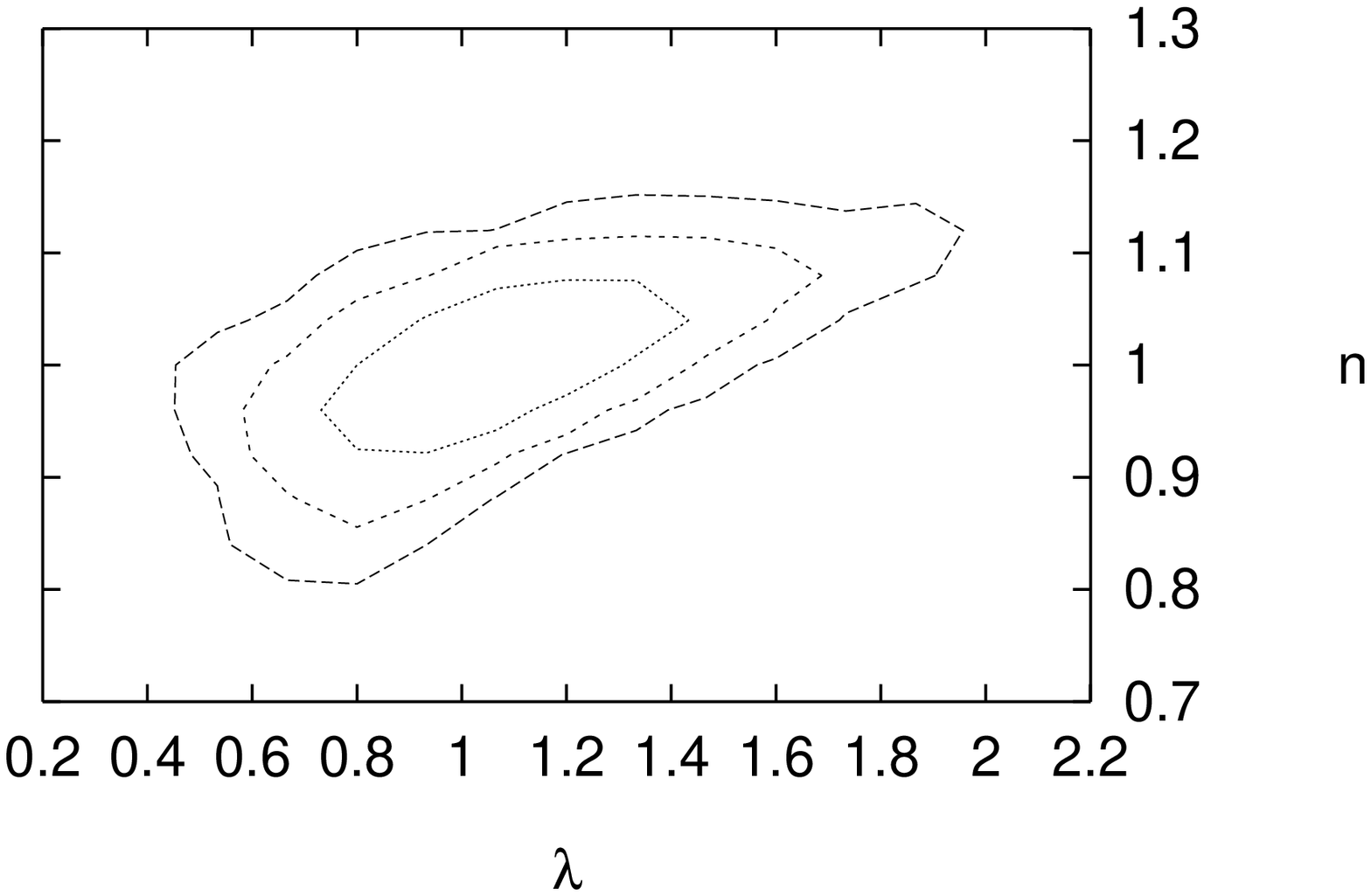,width=7cm,angle=-90} \hspace{-3cm}
\epsfig{file=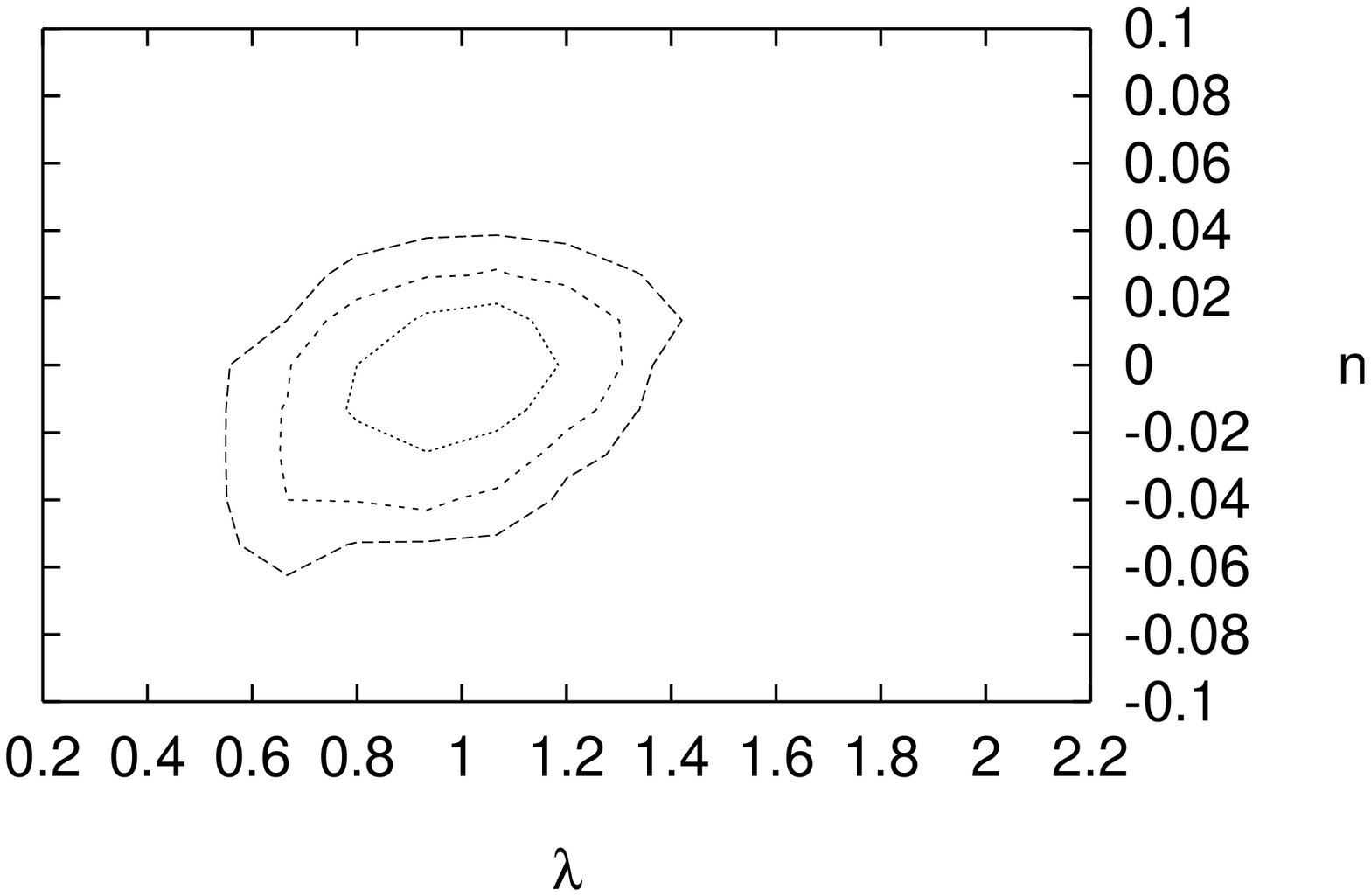,width=7cm,angle=-90}
\caption{We plot the $68.3,95.4 \text{ and } 99.7\%$ contour lines of $10^5$ cosmological models of our Markov chains. The upper two plots show a projection onto the $\lambda-\alpha$-plane; the left plots are for temperature and the right ones include polarization information. The lower two plots show the degeneracy of $\lambda$ with the slope of the spectrum of scalar perturbations, $n_S$. It can be seen in the upper and lower plots that polarization information reduces the degeneracy with the shape of the initial power spectrum.\label{mapcontours}}
\end{figure*}

In Section \ref{lambda} we discussed the different response to a
change of $\lambda$ of the temperature and the polarization
anisotropy. We found that on large angular scales polarization power
is increased if we increase the expansion rate during recombination,
while the temperature anisotropy is almost not affected on these
scales. When only temperature is being measured, changes in the
expansion rate of the universe during recombination are strongly
degenerate with the slope of the spectrum of the initial scalar
perturbations, which can be clearly seen in Figures
\ref{maphist} and \ref{mapcontours}. 
Because polarization responds differently on
different scales to a change of $\lambda$, it breaks this
degeneracy. We show this in Figure \ref{mapcontours}. The result of
the likelihood analysis which includes polarization information is
included in Table \ref{maplitem}.

We noticed that a major part of the information on $\lambda$ from MAP's polarization measurements will come from the cross correlation between temperature and polarization. When adding just the cross correlation information we found that we gain almost 90 \% of the information that is gained in the case in which all three estimators are included.

\begin{table*}
\begin{tabular}{|c|c|c|c|c|c|}
\hline
  T           & $(F_{ii}^{-1})^{1/2}$ &  $(F_{ii}^{-1})^{1/2}$, $\alpha$  &    T+P          & $(F_{ii}^{-1})^{1/2}$     &  $(F_{ii}^{-1})^{1/2}$, $\alpha$       \\
\hline   
$\lambda$    &     0.1928 (0.1279)    &    0.2616 (0.1992)   &  $\lambda$    &     0.1373 (0.0903)    &    0.1658 (0.1165)    \\    
\hline 
$n$          &   0.0294 (0.0248)          &     0.0646 (0.0595)     &   $n$             &    0.0170 (0.0144)   &      0.0489 (0.0418)           \\
\hline
 $\alpha$    & $\times$           &    0.0237 (0.0222)    &   $\alpha$      &     $\times$              &       0.0151 (0.0137)           \\
\hline

\end{tabular}
\caption{Fisher matrix results for MAP's expected 2 year data (in brackets are the values expected for a 4 year observation). The first two columns are the results when polarization information is not included. In the second column the curvature of the primordial spectrum was left to vary. In the last two columns  polarization information was included.\label{maplitem}}
\end{table*}

\subsubsection{Expected constraints from Planck}

\begin{table*}
\begin{center}
\begin{tabular}{|c|c|c|c|c|c|}
\hline
T  & $(F_{ii}^{-1})^{1/2}$ & $(F_{ii}^{-1})^{1/2}$, $\alpha$  &     T+P      &  $(F_{ii}^{-1})^{1/2}$ & $(F_{ii} ^{-1})^{1/2}$, $\alpha$   \\
\hline   
$\lambda$    &   0.0170 (0.0152)     &     0.0325 (0.0278)   & $\lambda$    &  0.0115 (0.0093)  &     0.0174 (0.0141)   \\    
\hline
$n$          &     0.0118 (0.0106)    &  0.0182  (0.0160)     & $n$       &     0.0072 (0.0060)   &     0.0098 (0.0080)   \\
\hline
$\alpha$      &      $\times$     &   0.0072  (0.0620)      &     $\alpha$       &    $\times$     &   0.0039  (0.0033)      \\
\hline
\end{tabular}
\end{center}
\caption{Fisher matrix results for Planck's estimated 1 year data (brackets contain the 2 year results). Columns are equivalent to those in Table \ref{maplitem}.\label{plancklitem}}
\end{table*}

\begin{figure*}
\epsfig{file=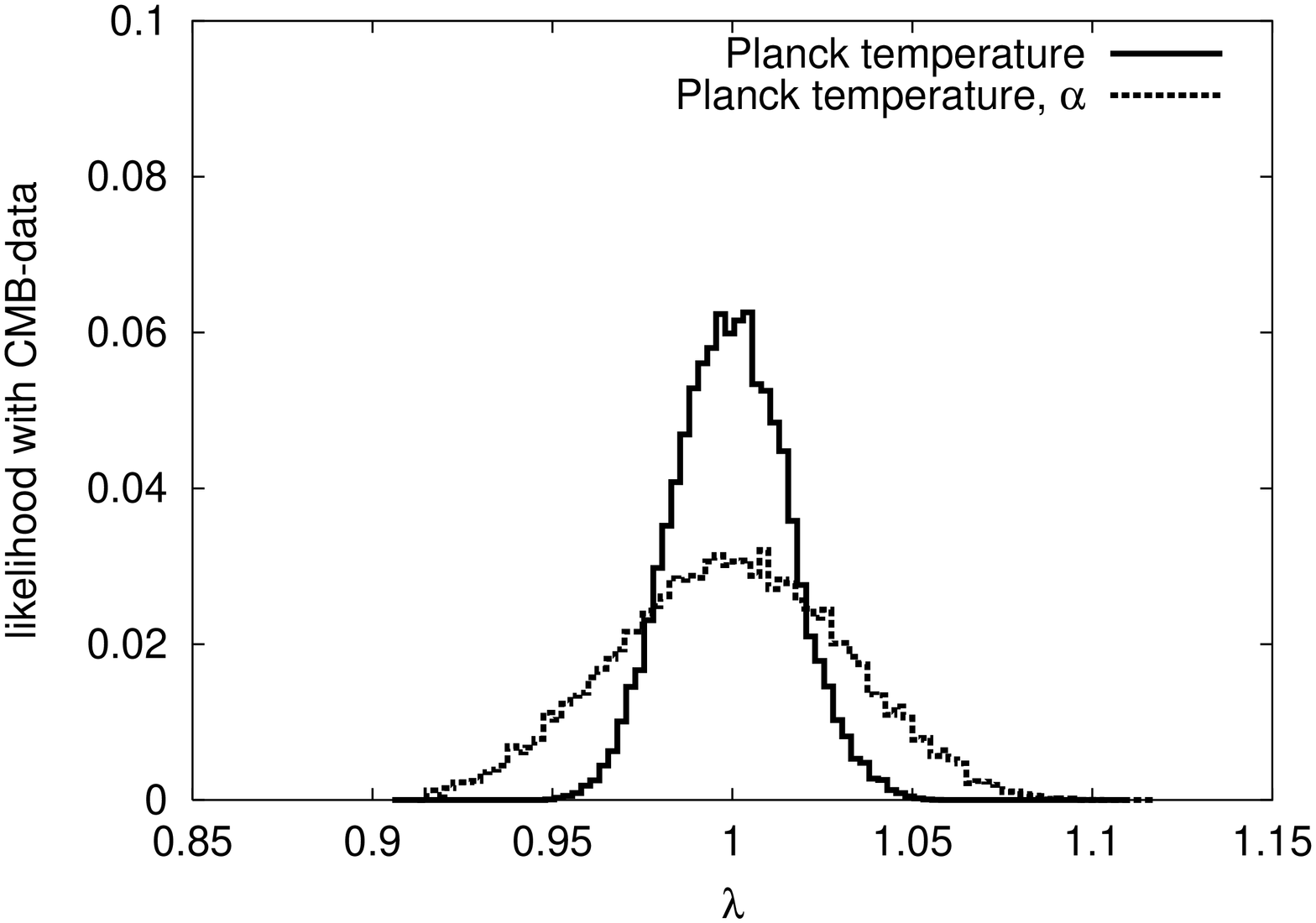,width=5cm,angle=-90} \hspace{1cm}
\epsfig{file=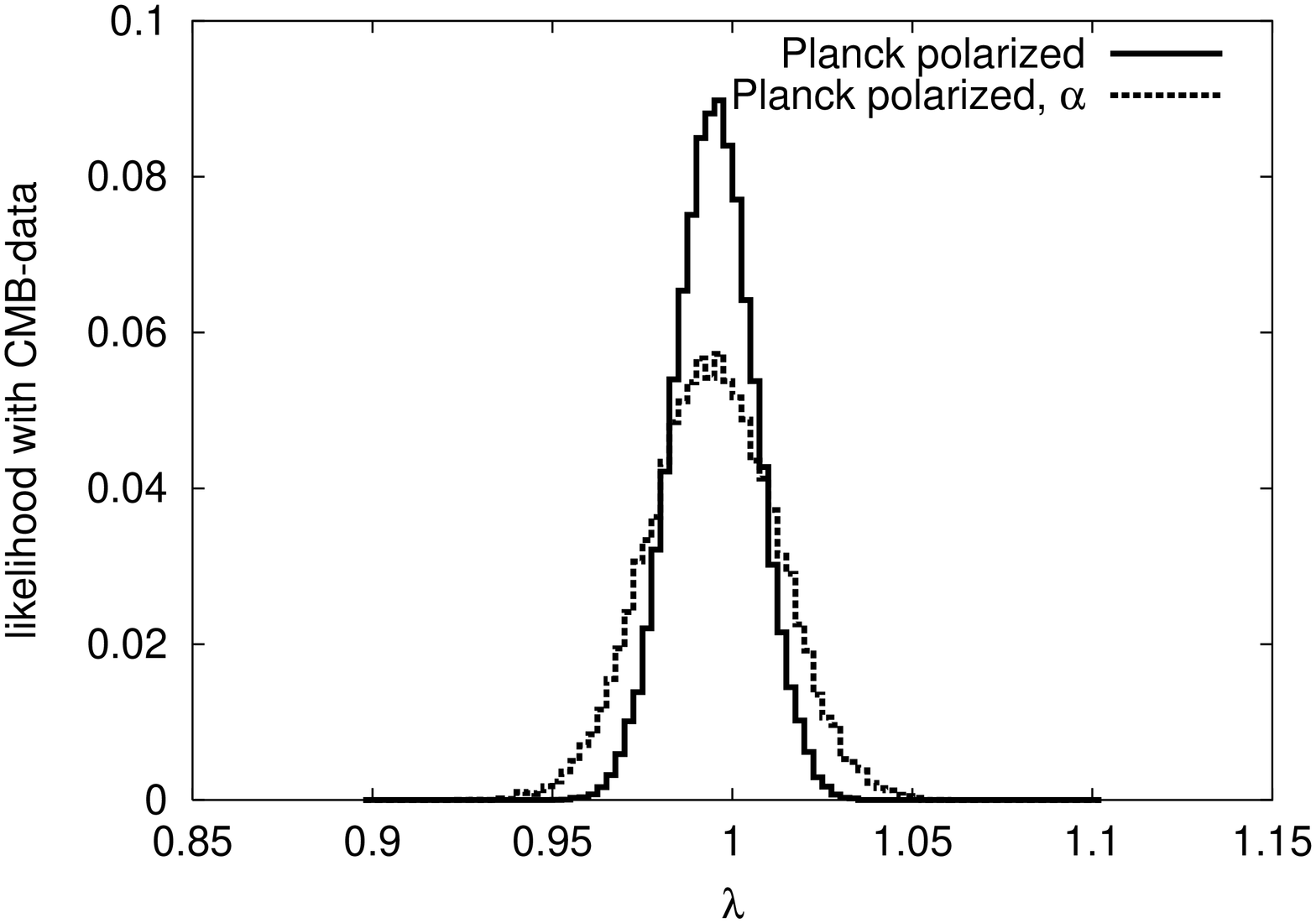,width=5cm,angle=-90}  
\caption{The distribution of values of $\lambda$ for one versus two degrees of freedom in the primordial power spectrum. The left plot shows the temperature case while the right includes polarization.\label{planckhist}}
\end{figure*}

\begin{figure*}
\epsfig{file=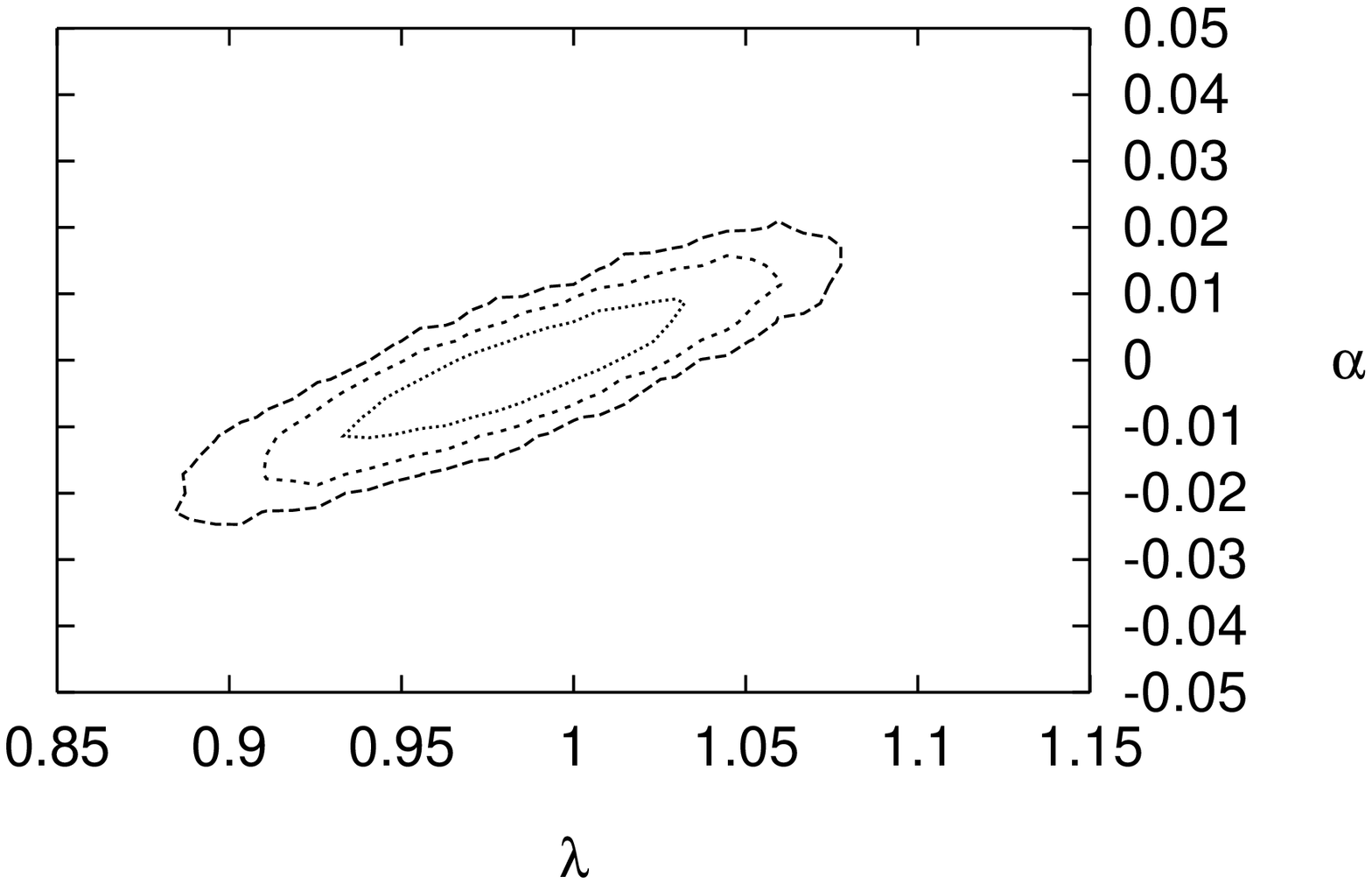,width=7cm,angle=-90} \hspace{-3cm}
\epsfig{file=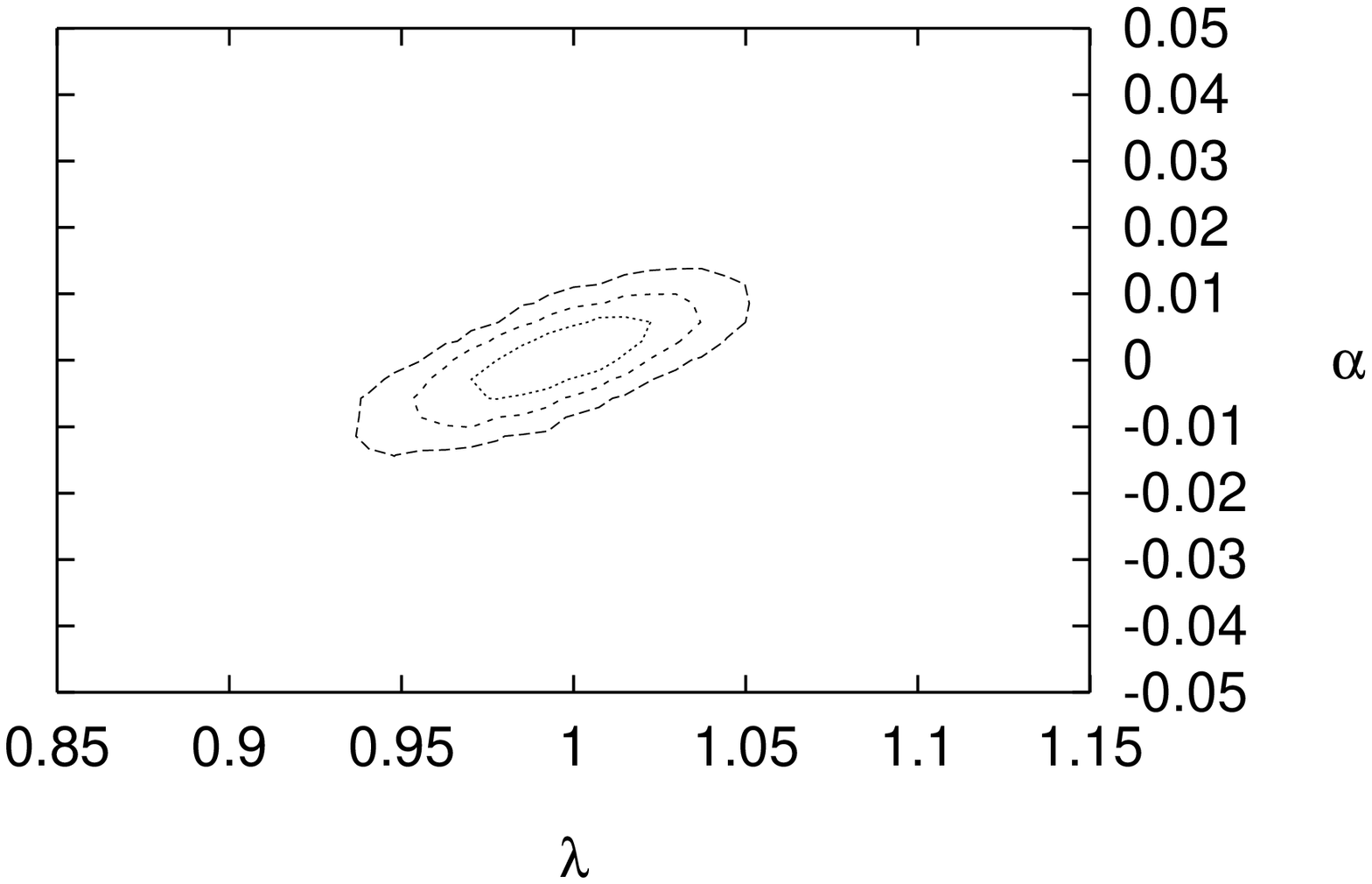,width=7cm,angle=-90}
\epsfig{file=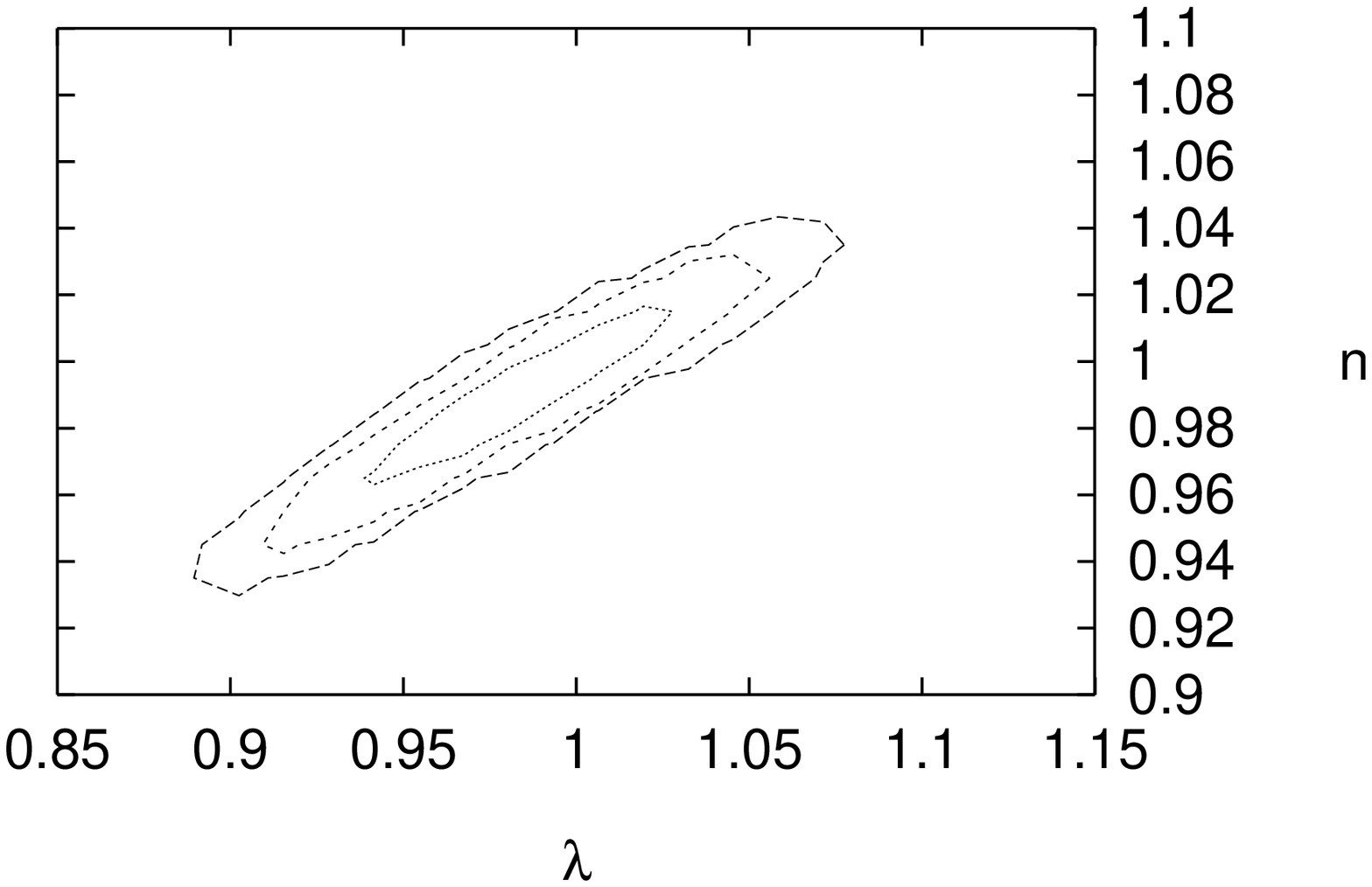,width=7cm,angle=-90} \hspace{-3cm}
\epsfig{file=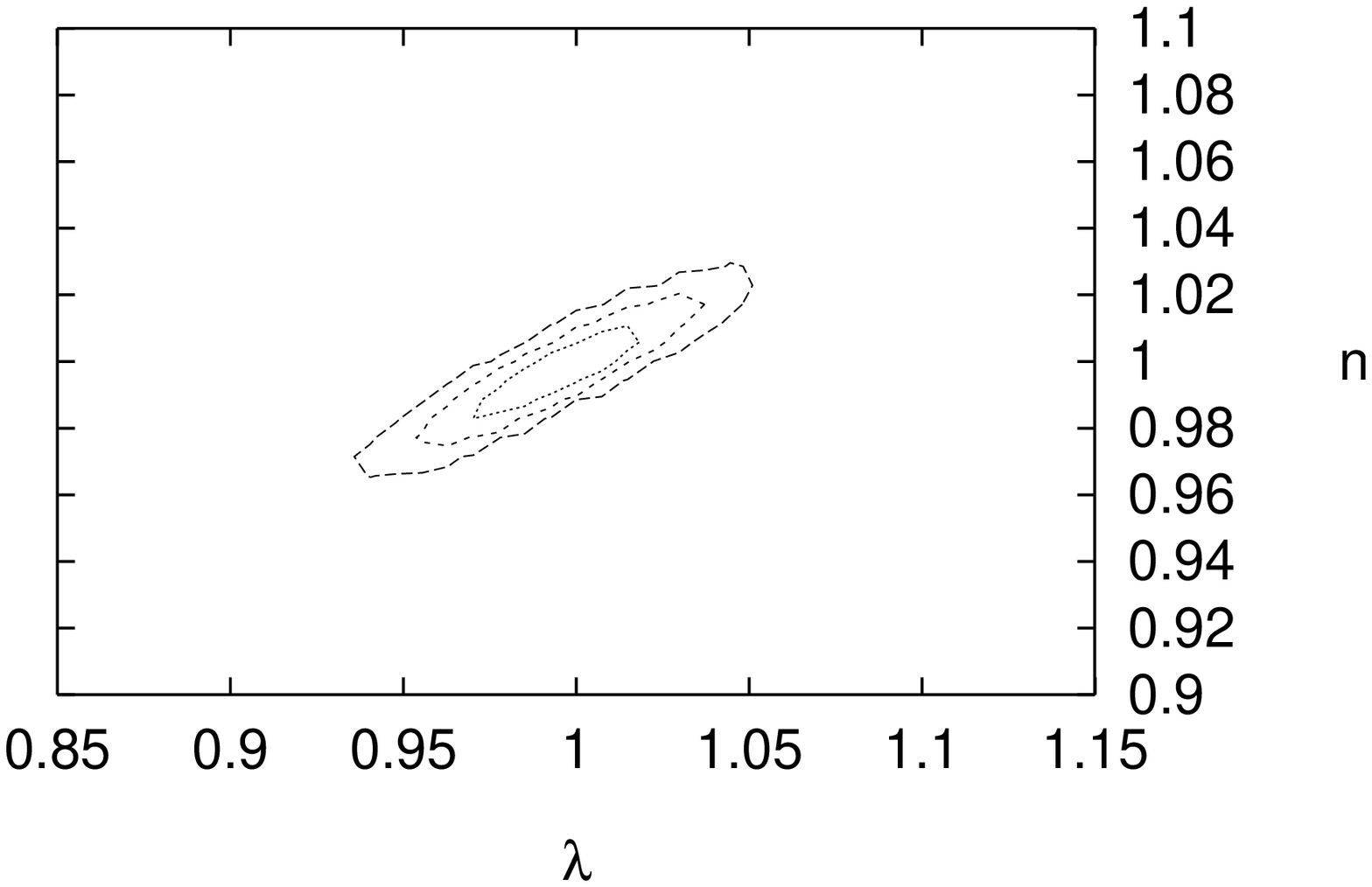,width=7cm,angle=-90}
\caption{Contour lines for the case of Planck (with the same ordering as above). Again it is shown that one can lift the degeneracy between $\lambda$ and the shape of the primordial fluctuations by measuring CMB polarization.\label{planckcont}}
\end{figure*}

The Planck satellite explores the very small structures in the primeaval plasma in a multipole range up to nearly $l=3000$. Even if only Planck's temperature data are used, the standard error in $\lambda$ will after a one year observation be as small as $3.2\%$, even if one marginalizes over the parameters that describe the shape of the primordial power spectrum (Table \ref{plancklitem}). This corresponds to a constraint on the gravitational constant of $\frac{\delta G}{G}=0.064$.
Finally, the improvement gained from Planck's polarization data ($\frac{\delta \lambda}{\lambda} \simeq 0.017 \Leftrightarrow \frac{\delta G}{G} \simeq 0.034$) is shown on the right hand side of that same Table. If one assumes that there is no curvature in the primordial power spectrum the constraint on G would be as small as $\frac{\delta G}{G} = 1.8 \%$ after an observation of two years. Figure \ref{planckhist} and \ref{planckcont} show again, analogously to the case of MAP, how polarization helps break the degeneracy between $\lambda$ and the parameters that describe the shape of primodial perturbations.

We have extended our investigation beyond Planck to the case of an
experiment which has essentially no noise and where the errors in the
CMB power spectra are on all scales dominated by the cosmic variance
term. For example experiments currently being considered for measuring
the B-modes of CMB polarization would be cosmic variance limited for E
polarization over a wide range of $ls$ (e.g. \cite{bmodes}). Such an optimal
experiment, exploring structures into a multipole range of $l=4000$,
represents the limit of how much information on $\lambda$ one could
extract from the CMB in principle. We found an expected error for
$\lambda$ of order $0.3\%$ which translates into a constraint of the
value of the gravitational constant during recombination of
$\frac{\delta G}{G} \simeq 0.6 \%$.

\section{Discussion\label{discussion}}

We have made the expansion rate of the universe a free parameter in a
likelihood analysis within an eight-dimensional cosmological parameter
space. For simplicity we assumed that the gravitational constant is
changed by a factor $\lambda$.  We showed that an increase of
$\lambda$ leads to a wider visibility function which in turn increases
the damping of anisotropies on small scales and increases the level of
large scale polarization.

We calculated the constraints that current CMB data can impose on the
expansion rate.  The constraints that can be imposed on the parameter
$\lambda$ using the information from the damping tail are severely
weakened by our lack of knowledge about the shape of the primordial power spectrum. 
We showed that measuring polarization helps to break this
degeneracy.  Current data can only constrain $\lambda$ to about 47\%
at 1 $\sigma$.  We showed that MAP could obtain 9\% error bars for
$\lambda$ while for Planck errorbars go down to 0.9\%. We also explored
the ultimate limit that could be achieved by a cosmic variance limited
experiment measuring anisotropies up to $l=4000$ and found errors of
under a percent in that case. Thus next generation
experiments should be able to deliver very accurate constraints on the
expansion rate of the universe during recombination.

We acknowledge that if the variation of the gravitational constant
during recombination is taken seriously a model needs to be built
where $G$ changes after recombination and converges towards the stable
value observed in laboratory experiments today and where its current
rate of change is less than the experimental bound $\frac{\dot{G}}{G}
\simeq 10^{-12} \text{yr}^{-1}$\cite{Uzan:2002vq,Guenther:1998}. If we introduce a scalar field
to control the value of $G$ we would also have to require that this
field does not lead to an unacceptably large fifth force and does not
violate solar system constraints such as shifting the orbit of the
moon through the Nordvedt effect \cite{Will,Nordtvedt}.

The shift of $G$ after recombination will induce a change in the angular
diameter distance to recombination, shifting the CMB power spectrum in
$l$. We have shown that future experiments will be able to constrain
the change of $G$ to a few percent due to its effect at
recombination. As a result the induced shift in the peak positions
would be small and could be interpreted as slightly different values
of $\Omega_m$ and/or $\Omega_\Lambda$, the two parameters that control
the distance to the last scattering surface in conventional models. In
the same way, any induced integrated Sachs-Wolfe (ISW) effect would be
difficult to observe because of the cosmic variance limitation. If
observed with a high enough accuracy it should not be identical with
what is predicted by a simple cosmological constant model.

In this paper we have shown that future measurements of the CMB
anisotropy will be able to extract information about the relation
between the expansion rate and the energy density of the universe
during recombination, because of its effect on the recombination
history of hydrogen.

\section{Acknowledgements}

OZ thanks Christian Armendariz-Picon, Hans-Joachim Drescher and Emiliano Sefusatti for useful discussion. OZ is supported by the Deutscher Akademischer Austauschdienst. MZ and OZ are supported by NSF grants AST 0098606 and PHY 0116590 and by the David and Lucille Packard Foundation Fellowship for Science and Engineering.



\end{document}